\documentstyle[multicol,aps, prl]{revtex}

\date{\today}
\draft
\tighten
\begin{document}
\def\sqr#1#2{{\vcenter{\hrule height.3pt
      \hbox{\vrule width.3pt height#2pt  \kern#1pt
         \vrule width.3pt}  \hrule height.3pt}}}
\def\square{\mathchoice{\sqr67\,}{\sqr67\,}\sqr{3}{3.5}\sqr{3}{3.5}}
\def\today{\ifcase\month\or
  January\or February\or March\or April\or May\or June\or July\or
  August\or September\or October\or November\or December\fi
  \space\number\day, \number\year}

\def\Bbb{\bf}


\newcommand{\ww}{\mbox{\tiny $\wedge$}}
\newcommand{\pp}{\partial}

\title{Branes, AdS gravitons and Virasoro symmetry\thanks{Preprint
numbers:  DAMTP-1999-164; \ LPTENS.99/47; CTP preprint \# 2925 }}
\author{ M Ba\~nados$^1$, A Chamblin$^2$ and G W Gibbons
$^3$\thanks{Permanent address DAMTP, University of Cambridge}}
\address{$^1$Departamento de F\'{\i}sica Te\'orica, Universidad de
 Zaragoza, Ciudad Universitaria, \\ Zaragoza 50009, Spain.\\
  $^2$Center for Theoretical Physics, MIT, Bldg. 6-304,
77 Massachusetts Ave.,\\ Cambridge, MA 02139, USA \\
$^3$Laboratoire de Physique Th\'eorique de l'Ecole Normale
Sup\'erieure \thanks{ Unit\'e Mixte de Recherche de Centre National de
la Recherche Scientifique et de l'Ecole Normale Superi\'eure }, 24 Rue
Lhomond, \\ 75231 Paris Cedex 05, France}
\maketitle

\begin{abstract}
We consider travelling waves propagating on the anti-de Sitter (AdS) 
background. It is pointed out that for any dimension $d$, this space
of solutions has a Virasoro symmetry with a non-zero central charge.
This result is a natural generalization to higher dimensions of the
three-dimensional Brown-Henneaux symmetry.  \\
\\
\end{abstract}

\begin{multicols} {2}

We consider in this note travelling waves in $d=3+n$ dimensions
propagating on AdS spacetime described by the line element,   
\begin{equation}
ds^2 = {l^2 \over z^2} [ dz^2 + H(z,x^i,u) du^2 + dudv + dx^i dx_i].  
\label{metric}
\end{equation}
Here $i=1,2,...,n$ and all coordinates are dimensionless. This metric
satisfies Einstein's equations with a cosmological constant
proportional to $-1/l^2$, provided $H$ satisfies the Siklos
\cite{GWG,Siklos85,GWGPJR} equation,
\begin{equation}
z^{d-2} \pp_z \left({1 \over z^{d-2}}\pp_z H\right)+ \nabla^2_x H=0.  
\label{Siklos}
\end{equation}
The solution (\ref{metric}) can also be obtained, for example,
following the general technique developed  in \cite{Garfinkle-V}
using anti-de Sitter 
space as background geometry. The perturbation $H\,
du^2/z^2$ is of the general Kerr-Shild type $\Phi(x) \xi_\mu \xi_\nu
\, dx^\mu dx^\nu$ where $\xi^\mu$ is a null, hypersurface orthogonal,
Killing field.  The dependence of $H$ in the coordinate $u$ is
arbitrary and defines the profile of the travelling wave. 

Solutions of the form (\ref{metric}), and their generalizations
including other gauge fields, have been extensively studied in the
string theory literature
\cite{Horowitz-T,Callan-MP,Dabholkar-GHW,Horowitz-M}. Recently,
(\ref{metric}) has appeared in \cite{Chamblin-G} as an exact
non-linear version of the Randall and Sundrum model \cite{Randall-S}.

In the particular case with $d=3$ and no transverse dimensions
($n=0$), the metric (\ref{metric}) enjoys a conformal symmetry which
follows from the analysis of \cite{BH}. There exists a transformation
of coordinates (non vanishing at infinity) whose associated
Noether
charges obeys the Virasoro algebra with a central charge $c=3l/2G_3$.
Actually, in three dimensions, the solution (\ref{metric}) can be
generalised to include an arbitrary function of $v$ (see \cite{Baires}
for the explicit form of the metric). This yields two copies of the
Virasoro algebra with the same central charge, and hence, the
conformal group.  This result has provided an elegant statistical
description of (non-extreme) black hole entropy in three dimensions
\cite{Strominger97}. 

The goal of this note is to point out that the results of \cite{BH}
can be extended in a natural way to higher dimensions. Indeed, we
shall prove that the {class of metrics}  (\ref{metric}) has a Virasoro
symmetry
with a central charge
\begin{equation}
c = {3l^{d-2} \over 2G_d},
\label{c}
\end{equation}
where $G_d$ is the $d$-dimensional Newton constant.  

The existence of an infinite dimensional symmetry $u \rightarrow
f(u)$ is signaled by the arbitrarness of $H$ as a function of $u$.
This, together with the adS structure, yields the Virasoro algebra. In
fact, the line element (\ref{metric}) is form-invariant under
the following redefinition of the coordinates,
\begin{eqnarray}
 u   &=& f(u') \nonumber\\
 x^i &=& x'^i \sqrt{\pp' f}\nonumber \\
 z   &=& z' \sqrt{\pp' f}  \nonumber \\
 v   &=& v' - {1 \over 2} (x'^ix'_i + z'^2) {\pp'^2 f \over \pp' f }
\label{trans} 
\end{eqnarray}
where $f$ is an arbitrary function of its argument. This
transformation was already known to Siklos \cite{Siklos85}. In the
primed coordinates, the metric has the same form as (\ref{metric})
with 
\begin{equation}
H'(z',x',u') = H(z,x,u) \pp' f - {1 \over 2} (z'^2 + x'^ix'_i)
\{f,u'\},
\label{H'H}
\end{equation}
where $\{f,u\}$ denotes the Schwarzian derivative of the map $f$,
\begin{equation}
\{f,u\} = {\pp^3 f \over \pp f} - {3 \over 2} \left({\pp^2 f \over \pp
f}\right)^2.
\end{equation}

The transformation (\ref{trans}) is not a Killing symmetry of
(\ref{metric}) because it changes the metric. However it does
take solutions to solutions. In other words it acts on the
infinite-dimensional simplectic manifold consisting of classical
solutions of Einstein's Equations of the form (\ref{metric}). When a
group acts on a simplectic manifold preserving  the symplectic form the
associated Poisson algebra will in general be a central extension of
the Lie algebra of the original group. In our case the group is that
of one-dimensional diffeomorphisms and the central extension that of
Virasoro.   
 
To make contact with previous work we note that the transformation
only changes the value of $H$ leaving the leading part of the metric
invariant, it defines an asymptotic symmetry. Generically speaking, a
transformation of coordinates is trivial if it goes to the identity
sufficiently fast at infinity. In our choice of coordinates, infinity
is located in the open region\footnote{Note, however, that
(\ref{metric}) is not globally asymptotically AdS.} $z\rightarrow 0$
and since the
map $u \rightarrow f(u)$ does not depend on $z$, the transformation
(\ref{trans}) does not go to the identity there. In fact, the
transformation (\ref{trans}) is an asymptotic symmetry of the space of
solutions (\ref{metric}) with a corresponding non-zero Noether charge.

By a straightforward application of the results presented in 
\cite{Regge-T,BH} and references therein, the Noether charge
associated to (\ref{trans}) is\footnote{We use the
coordinate $v$ as time, and we assume that $u$ is compact with period
$2\pi$, as in DLCGQ. See, for example, \cite{Susskind}. This yields a
meaningful ADM decomposition provided $H\neq 0$.},  
\begin{equation}
J(\epsilon) = \lim_{z\rightarrow 0}  \int  { d^n x \over z^n}\,
\int {du\over 2\pi} \,\epsilon(u) \, T(z,x^i,u).
\label{J}
\end{equation}
Here $\epsilon(u) = f(u)-u$ is the infinitesimal parameter of the
transformation, and $T$ is related to $H$ by,
\begin{equation}
T(z,x^i,u) = {l^{d-2} \over 8G_d }\, {1 \over z} \pp_z H(z,x^i,u).
\label{TH}
\end{equation}
In the calculation of $J$ we have assumed that the metric at infinity
($z \rightarrow 0$) takes the form (\ref{metric}) for arbitrary values
of $H$. 

The relevant properties of the transformation (\ref{trans}) are
encoded in the Noether charge $J$. If $J$ is finite, then the
transformation (\ref{trans}) represent a physical --not gauge-- change
of the state. This is analogous to a boost in an asymptotically flat
space which induces momentum. Furthermore, the knowledge of the
charge $J$ and its transformation properties will enable us to
determine the algebra of the generators of (\ref{trans}) in a
straightforward way.     

The function $T$ appearing in (\ref{J}) transforms as a Virasoro
operator. Indeed, from (\ref{H'H}) and (\ref{TH}) it is
straightforward to prove (applying the operator ${1 \over z'} \pp_{z'}
= (\pp' f) {1 \over z} \pp_z $ at both sides of (\ref{H'H})) that
$T$ transforms as,  
\begin{equation}
T'(z',x',u') = T(z,x,u) (\pp' f)^2 - {c \over 12} \{f,u'\},
\label{T'T}
\end{equation}
where the central charge is given in (\ref{c}). 

Let us now analyse the value of the Noether charge $J$ on the space of
solutions of the equation (\ref{Siklos}). The general solution to
(\ref{Siklos}) can be written down and involves Bessel functions.
Actually, the function $T$, related to $H$ in (\ref{TH}), satisfies a
Bessel equation as well.  Since in our discussion $H$ is not relevant
(the charge and Virasoro transformation only depend on $T$
\footnote{The components of the curvature tensor in an orthonormal
frame, which control the existence of singularities \cite{Chamblin-G},
depend only on $T$ as well.}) we write the general solution directly
in terms of $T$, 
\begin{eqnarray}
T(u,z,x^i) &=& \int d^n k \, z^{n/2} [ A(k_i,u) J_{n/2}(kz)
\nonumber \\ &&\ \ \ \ \ \ + \  B(k_i,u) Y_{n/2} (kz) ] e^{ik_i x^i }, 
\label{T}
\end{eqnarray}
where $k=\sqrt{k_ik^i}$, $J_{n/2}$ and $Y_{n/2}$ are Bessel
functions, and $A$ and $B$ are arbitrary functions of their arguments.  

Inserting (\ref{T}) into (\ref{J}) we find the formula for
the Noether charge,  
\begin{equation}
J(\epsilon) = \lim_{z\rightarrow 0} \int {du\over 2\pi} \, \epsilon(u)
\left[A(0,u) + {1 \over z^n} B(0,u)\right],
\label{J1}
\end{equation}
showing that, as usual, it only depends on the zero modes.  In this
formula we have absorbed some irrelevant coefficients into
redefinitions of $A$ and $B$. 

The sector  $(A\neq 0,B=0)$ has a finite Noether charge in the limit
$z \rightarrow 0$. Note, however, that the central term in (\ref{T'T})
does not depend on $z$. Since the solution with $A\neq 0$ has a
prefactor $z^{n/2} J_{n/2}(kz) \sim z^n$ we conclude that the central
piece in (\ref{T'T}) is not carried by this sector. Since we are
interested in the centrally extended Virasoro algebra we set, from now
on, $A=0$.  

Next, we note that the central piece in (\ref{T'T}) does not depend on
$x^i$ either. This means that only the zero mode of $B(k_i,u)$ will
see it.  Setting $B(k_i,u) = B(u)$ we find the formula for the charge
\begin{equation}
J_B(\epsilon) = V_\perp  \int {du \over 2\pi }
\epsilon(u) B(u),
\end{equation}
where the prefactor $V_\perp:= \int d^n x / z^n$ is the
proper volume of  transverse space (recall that $z^{-n}$ is the factor
in the adS volume element).  Furthermore,  from (\ref{T'T}) we derive
the transformation law of $B(u)$, 
\begin{equation}
B'(u') =  B(u) (\pp' f)^2 - {c \over 12} \{f,u'\}.
\label{B'B}
\end{equation}

As usual when working with extended objects, the charge $J_B$ diverges
because $V_\perp \rightarrow \infty$. We then define the density of
charge per unit of proper volume,
\begin{equation}
j_B(\epsilon) = {J_B(\epsilon) \over V_\perp} = \int
{du \over 2\pi } \epsilon(u) B(u),
\label{j}
\end{equation}
which is indeed finite.    

The formula (\ref{B'B}) is not enough to claim that $B$ carries a
central charge $c$ because it depends on the normalization of $B$. In
order to fix the value of the central charge we need to prove that $B$
satisfies the Virasoro algebra.  This can be done using a general
theorem proved in \cite{Regge-T,BH,BH2}. Let $(M,g)$ be a Riemannan
spacetime with a group of asymptotic Killing vectors fields
$\xi^\mu_i$ ($i=1,...,N$), and let
$H[\xi_i]=\int \xi^\mu_i {\cal H}_\mu + J[\xi_i]$ the corresponding
canonical generators.  By definition, $J[\xi_i]$ is a boundary term
that makes $H$ differentiable and is equal to the Noether charge
associated to the symmetry $\xi_i$ \cite{Regge-T}. Then, it follows,  
\begin{equation}
\delta_{\xi_i} J(\xi_j) = [J(\xi_i),J(\xi_j)] .
\label{PB}
\end{equation}
This formula yields a powerful way to compute the algebra of global
charges in gauge theories. We remark that the number of asymptotic
symmetries need not to be finite. Indeed, in our case, as in
\cite{BH}, the asymptotic symmetries are parametrised by the infinite
number of Fourier modes of $f(u)$. Another example on which this
procedure yields an infinite dimensional algebra is the affine
Kac-Moody algebra in non-Abelian Chern-Simons theory \cite{B}.     

We now apply this result to our situation. First, we notice that since
we are working with a density of charge, the right hand side of
(\ref{PB}) should be understood as a density of Poisson bracket, i.e.,
as the Poisson bracket divided by the proper volume $V_\perp$. The
function $B(u)$ transforms as in (\ref{B'B}). We then find, 
\begin{eqnarray}
&& \int {du \over 2\pi }\epsilon \left( \gamma \pp B + 2 B\pp\gamma -
{c\over 12} \pp^3 \gamma \right) \nonumber\\ && \ \ \ = \  
\int {du \over 2\pi }  \int {du' \over 2\pi } \epsilon(u)
\gamma(u')[B(u),B(u')], 
\end{eqnarray}
which is indeed the Virasoro algebra with a central charge $c$.

The Virasoro symmetries described here arise generically for all
solutions describing travelling waves on adS backgrounds, or whose
near horizon geometries are of the form $adS_n\times N_{d-n}$. These
include many examples relevant for the adS/CFT correspondence as, for
example, D3 branes and the M2/M5 solution. Consider, in particular,
the D1/D5 system consisting of a ten dimensional string theory
compactified on $S_1\times T_4$ with a non-zero RR field $H_3$
carrying both electric and magnetic charges $Q_1$ and $Q_5$. This
system can be described by a two dimensional conformal field theory
with a central charge $c=6Q_1Q_5$ \cite{Strominger-V}. On the other
hand, the near horizon geometry has the well-known structure $adS_3
\times S_3 \times T^4$ and we should then expect a realisation of the
Brown-Henneaux conformal symmetry \cite{BH}. It was noticed in
\cite{Strominger97} that the effective three-dimensional Newton
constant $G_3$ and cosmological constant $l$ satisfy $c=3l/2G_3=6Q_1
Q_5$. Thus, the Brown-Henneaux Virasoro generators are indeed those of
the CFT.  

Travelling waves propagating on this background have been studied in
\cite{Horowitz-M}. It was already noticed in that reference that there
exists a transformation of coordinates mapping the longitudinal
travelling wave into the background geometry. This transformation is a
particular case of the general Virasoro symmetry discussed here and
corresponds to solving the differential equation $T'(u')=T(u) (\pp'
f)^2 - {c\over 12} \{f,u'\}=0$.  Note that a function $f(u)$ mapping
$T\neq 0$ into $T'=0$ can exist because of the inhomogenous term in
the transformation law of $T$. Let us now check that the central
charge, appearing in the classical Poisson bracket of the Noether
charges, is the correct one.   

The (extreme) ten dimensional metric describing a longitudinal
travelling wave is (see \cite{Horowitz-M} and references therein)
\begin{eqnarray}
 ds^2 &=& \left( 1+ {r_0^2 \over r^2 } \right)^{-1} \left[dudv +
 {p(u) \over r^2} du^2 \right] \nonumber \\ && \ \ \ + \ \left( 1+
 {r_0^2 \over r^2 }
\right)(dr^2 + r^2d\Omega_3)
+ dy_i dy^i 
\label{d1d5}
\end{eqnarray}
where $y^i$ are coordinates on $T^4$, whose volume is $(2\pi)^4V$. The
electric and magnetic charges are given by 
\begin{equation}
Q_1 = {V\, r_0^2 \over g}, \ \ \ \ Q_5 = {r^2_0 \over g} ,
\end{equation}
where $g$ is the string coupling related to Newton's constant as
$G_{10}=8\pi^6 g^2$. The near the horizon geometry ($r \sim 0$) is
$adS_3 \times S_3 \times T^4$ and we can then apply the
transformations (\ref{trans}) adapted to these coordinates. The
resulting Noether charge is 
\begin{equation}
J(\epsilon) = {1 \over 16\pi G_{10}} \int d^4y \int r_0^3 d\Omega_3
\int r_0 du \, \epsilon(u)\, 2 p(u) .
\end{equation}
In this case the volume of transverse space is finite, in fact the
problem is effectively three-dimensional. Inserting the values for
$G_{10}$, $V$ and $r_0$ we find
\begin{equation}
 J(\epsilon) = Q_1 Q_5 \int {du\over 2\pi} \epsilon(u) p(u)
\end{equation}
and the combination $T := Q_1 Q_5\, p(u)$ satisfies the Virasoro
algebra with central charge $c=6Q_1Q_5$, as desired.

\acknowledgments

MB would like to thank Marc Henneaux for many useful conversations on
asymptotic properties of adS, and also Don Marolf for useful
correspondence. MB also thanks financial support from
CICYT (Spain) grant AEN-97-1680, and the Spanish postdoctoral program
of Ministerio de Educaci\'on y Cultura.   AC is supported in part 
by funds provided by the U.S. Department of Energy (D.O.E.) under
cooperative research agreement DE-FC02-94ER40818.

\end{multicols}

\end{document}